\documentclass[epj,final]{svjour} \usepackage{graphicx,amsmath,bm}

\begin{document}

\title{The quantum vacuum at the foundations of classical
  electrodynamics}

\author{
G. Leuchs\inst{1,2} 
\and A. S. Villar\inst{1,2} 
\and  L. L. S\'anchez-Soto\inst{1,3}} 

\institute{
Max Planck Institut f\"ur die Physik des Lichts, 
G\"unther-Scharowsky-Str. 1/Bau 24, Erlangen, Germany 
\and 
Institut f\"ur Optik, Information und Photonik,
Universit\"at Erlangen-N\"urnberg, Staudtstr. 7/B2, 91058 Erlangen, Germany 
\and 
Departamento de \'Optica, Facultad de F\'\i{}sica,
 Universidad Complutense, 28040 Madrid, Spain}
\mail{gerd.leuchs@mpl.mpg.de}

\date{\today}

\abstract{
  In the classical theory of electromagnetism, the permittivity
  $\varepsilon_0$ and the permeability $\mu_0$ of free space are
  constants whose magnitudes do not seem to possess any deeper
  physical meaning. By replacing the free space of classical physics
  with the quantum notion of the vacuum, we speculate that the values
  of the aforementioned constants could arise from the polarization
  and magnetization of virtual pairs in vacuum. A classical dispersion
  model with parameters determined by quantum and particle physics is
  employed to estimate their values. We find the correct orders of
  magnitude. Additionally, our simple assumptions yield an independent
  estimate for the number of charged elementary particles based on the
  known values of $\varepsilon_0$ and $\mu_0$ and for the volume of a
  virtual pair. Such interpretation would provide an intriguing
  connection between the celebrated theory of classical
  electromagnetism and the quantum theory in the weak field limit.
}
\maketitle

In classical electrodynamics, a dielectric medium becomes polarized in
the presence of an electric field. A weak electric field slightly
displaces the electronic clouds from their binding nuclei, producing
tiny atomic and molecular dipoles in the material. The macroscopic
effect, averaged over a region large compared to the atomic
dimensions, is the appearance of an induced electric field. The
strength of this response depends on how susceptible to displacement
the atomic dipoles are and on how much space they occupy.

Putting it in more quantitative terms, in an isotropic medium the
polarization $\vec P$ is proportional to the external electric field
$\vec E$ through the relation $\vec P=\chi\varepsilon_0\vec E$, where
$\chi$ is the linear susceptibility of the material and
$\varepsilon_0$ is the permittivity of free space. The constant
$\varepsilon_0$ has the unimpressive role of matching units, just
`happening to have the value'
$\varepsilon_0=8.8542\times10^{-12}$~As/(Vm)~\cite{griffiths}. The
expression for the magnitude of $\vec P$, which connects the
microscopic dipoles to their macroscopic effect, reads
\begin{equation}
  P = \frac{\wp}{V}\,,
  \label{defP}
\end{equation}
where $\wp$ is the dipole moment and $V$ is the effective volume per
dipole.

It is customary to define the electric displacement as
\begin{equation}
  \vec D = \varepsilon_0\vec E + \vec P \,,
  \label{defD}
\end{equation}
since it turns out to be independent of the induced charge
\cite{tamm}. In the particular case of an isotropic linear dielectric,
Eq.~(\ref{defD}) yields $\vec D=\varepsilon\varepsilon_0\vec E$, where
$\varepsilon=(1+\chi)$ is the relative permittivity of the medium.

The electric displacement $\vec D$ enjoys no better status among us
physicists than that of $\varepsilon_0$. It is mostly seen as a
mathematical tool of limited use, because it would combine two
physically different quantities~\cite{tamm}. In fact, the electric
field is the force per unit charge, whilst the spatial variation of
the polarization is the induced charge. Sometimes, the electric
displacement is even considered completely dispensable~\cite{purcell},
since the condition $\vec\nabla\times\vec D=0$ does not apply in
general --- in contrast to the case of $\vec E$, a scalar potential
cannot be associated with $\vec D$.

In this paper, we offer a physical interpretation of $\varepsilon_0$
and thus of $\vec D$. We discuss how Eq.~(\ref{defD}) might hint at
fundamental physical aspects of the electromagnetic field itself if,
in the light of its second term, we interpret the first term as also
originating from a polarized medium, so that $\vec D$ is the total
polarization. That would put $\varepsilon_0\vec E$ and $\vec P$ on an
equal footing\footnote{In Gaussian units, Eq.~(\ref{defD}) reads $\vec
  D = \vec E + 4\pi\vec P$, so that $\varepsilon_0$ is replaced by the
  constant `1.' In the language of Gaussian units, this paper
  discusses the physical significance of this `1.'}. That
interpretation is preposterous in classical electromagnetism, since
the vacuum is defined as the emptiness. In quantum theory, however,
that is not the case, and virtual particle pairs making up the vacuum
are expected to show a response to an applied field. In analogy to
$\varepsilon\varepsilon_0$ for a material medium, $\varepsilon_0$
itself could be associated with the polarization of the quantum vacuum
$P_0=\varepsilon_0E$, a concept deeply rooted in quantum
electrodynamics~\cite{milonni,gies}. We later extend the discussion to
magnetic phenomena, linking the permeability of free space $\mu_0$ to
the magnetization of virtual pairs in vacuum.

The presence of an external electromagnetic field disturbs the
stability of the vacuum as the ground state of all fields. In this
case, virtual particles must be considered to account for the possible
physical processes in vacuum. Phenomena such as pair creation, vacuum
birefringence, and light-light scattering are well established for
strong fields~\cite{heisenbergeuler,sauter1,sauter2,schwinger}, with
amplitudes larger than $E_S=m^2c^3/e\hbar\approx10^{18}$~V/m. This is
exactly the expression for the critical field first pointed out by
Niels Bohr, as acknowledged by F.~Sauter\footnote{See second footnote
  on page 743 of Ref.~\cite{sauter1}: ``I would like to thank
  Prof. Heisenberg for kindly informing me about this hypothesis of
  N. Bohr.''}. In the literature, $E_S$ is often referred to as the
Schwinger field~\cite{schwinger}. These effects illustrate the
nonlinear properties of the vacuum~\cite{milonni,gies}.

Here we assume the limit of weak fields and focus on the linear
response. The question we pose is: could the vacuum quantum
fluctuations be subtly embedded in the classical theory? If yes, one
piece of evidence could be the existence of physical constants whose
numerical values are simply determined experimentally, but which would
emerge naturally from the quantum theory. We speculate whether the
permittivity and the permeability of free space could be such
quantities.  In the remainder of this paper, we will estimate the
values of $\varepsilon_0$ and $\mu_0$ based on simple semi-classical
assumptions.

We start by considering the electrical properties of the quantum
vacuum. An external electric field interacting with the virtual
electron-positron pairs polarizes them. The magnitude of the dipole
moment $\wp=ex$ induced on one virtual pair (where $e$ is the electron
charge and $x$ is the displacement) can be computed by considering the
virtual pair as a harmonic oscillator, such that in the quasi-static
limit
\begin{equation}
  m\omega_0^2x=eE\,,
  \label{oscdisp}
\end{equation}
where $m$ is the electron mass and $\omega_0$ is the natural resonance
frequency. As customary in this sort of semi-classical treatment,
valid when the excitation of the quantum oscillator is very small (or
equivalently $E\ll E_S$), the resonance frequency is determined by the
energy associated with the quantum transition. The two-level system
under consideration is formed by the ground state of the virtual pairs
and the real positronium atom as the excited state~\cite{keitel}. The
energy gap is denoted $\mathcal{E}_\mathrm{gap}$, so that
$\omega_0=\mathcal{E}_\mathrm{gap}/\hbar$, and it is expected to be of
the order of the positronium rest energy. In addition, the
quasi-static regime requires that the frequency $\omega$ of the
classical field $E$ be much smaller than the oscillator resonance
frequency $\omega_0$. That means we are considering a low-energy field
probing pair creation as a non-resonant effect (classical
field). Inserting this relation for $\omega_0$ into
Eq.~(\ref{oscdisp}) to determine the displacement, one obtains the
induced dipole moment
\begin{equation}
  \wp=\frac{e^2}{m \omega_0^2}E\,.
  \label{dipmoment}
\end{equation}

The magnitude of the vacuum polarization given by Eq.~(\ref{defP})
depends on the effective volume per dipole $V=r^3$. In theoretical
investigations~\cite{narozhny}, it is generally accepted that the
appropriate length scale for a virtual electron-positron pair in
vacuum is the Compton wavelength\footnote{The uncertainty principle
  also offers an estimate for this quantity, by the condition $\Delta
  x \Delta p\geq \hbar/2$, where $\Delta x$ and $\Delta p$
  respectively denote the position and the momentum
  uncertainties. Taking the momentum $\Delta p=\hbar\omega_0/c$
  necessary to close the rest energy gap would yield $\Delta x\geq
  c/2\omega_0$, which is of the order of the Compton wavelength. The
  exact choice on how to relate $\Delta x$ to the dipole volume is
  arbitrary to a large extent when based on the uncertainty
  inequality.} $\lambda_c=\hbar/mc$.  With these considerations,
Eq.~(\ref{defP}) results in the vacuum polarization\footnote{The same
  result would be derived in Gaussian units, for which
  $\varepsilon_0=1/4\pi$.}
\begin{equation}
  P_0=\frac{e^2}{m\omega_0^2r^3}E
  \label{polariza}
\end{equation}
The induced polarization is proportional to the electric field, as
expected. Assigning the quantity multiplying $E$ to the vacuum
polarizability, we arrive at\footnote{Substituting typical numbers in
  Eq.~(\ref{varepsilonvalue}), for instance $\hbar\omega_0=2mc^2$ as
  the rest energy of the positronium and $r=\lambda_c/2$, one would
  get $\tilde\varepsilon_0=2e^2/\hbar c=1.62\times10
  ^{-12}$~As/(Vm). This number underestimates the correct value by one
  order of magnitude. However, apart from numerical prefactors which
  could possibly change by a small amount, we note that the expression
  for $\tilde\varepsilon_0$ does not depend on the rest mass if the
  energy gap is proportional to the latter. As discussed in what
  follows, this fact hints at having to sum the contributions from all
  elementary particles with electric charge, in this manner
  reinterpreting the above value of $\tilde\varepsilon_0$ as the
  partial contribution from the virtual electron-positron pairs. This
  approach provides an estimate for the number of elementary pairs
  contributing to the vacuum response.}
\begin{equation}
  \tilde\varepsilon_0 = \frac{e^2}{m\omega_0^2r^3}\,.
  \label{varepsilonvalue}
\end{equation}

For the permeability of free space $\mu_0$, we attempt a similar
procedure to estimate its value by associating this quantity to the
linear magnetic response of the quantum vacuum.
Here we encounter the added difficulty that some aspects of the
material magnetization arise from pure quantum effects without
classical explanation. One must resort in this case to crude
semi-classical assumptions that, nevertheless, have often proved
useful in determining the correct order of magnitude of magnetic
effects~\cite{feynman}. We will continue our analysis adapting results
from quantum mechanics to obtain $\mu_0$ from a seemingly naive
classical picture.

The magnetic counterpart of Eq.~(\ref{defD}) is
\begin{equation}
  \vec H = \frac{1}{\mu_0}\vec B-\vec M\,,
\end{equation}
where $\vec M$ is the magnetization and $\vec H$ is sometimes called
the `magnetic field,' although we will simply call it $\vec H$ and
reserve the term for $\vec B$. In analogy to $\vec D$, the vector
$\vec H$ isolates the contribution from the free current to the total
magnetic field. The expression above hints at associating $\mu_0$ to
the inverse of the vacuum magnetic response.

Analogously to the polarization, the magnitude of $\vec M$ can be
calculated from the microscopic magnetic dipole moment $\mathcal{M}$
and its effective volume using
\begin{equation}
  M = \frac{\mathcal{M}}{V}\,,
  \label{defM}
\end{equation}
where for consistency the same volume used for $\wp$ is assumed.

An external magnetic field applied to the vacuum induces an electric
field vortex which accelerates the virtual electron and positron in
opposite directions. This electric field $E_i$ induced in a circular
orbit fulfills
\begin{equation}
  E_i=- \frac{r}{2}\dot B\,,
\end{equation}
where the dot denotes the time derivative. It generates a torque that
increases the electron angular momentum $J$ by
\begin{equation}
  \Delta J=\frac{er^2}{2}B\,.
  \label{deltaJ}
\end{equation}

According to quantum mechanics, the magnetic moment and the angular
momentum relate to each other through the expression
\begin{equation}
  M=\frac{ge}{2m}J\,,
  \label{magnetJ}
\end{equation}
where $g$ is the Land\'e factor. A completely orbital or diamagnetic
contribution would render $g=1$, whilst $g=2$ would correspond to a
pure spin or paramagnetic contribution. We choose the latter
possibility for definiteness. Due to symmetry and simplicity, the
angular momentum state of the virtual pair is assumed a singlet, in
which case the positron contributes the same amount as the electron to
the magnetic response. These particles have thus opposite spins and
orbital angular momenta, but because of their opposite charges, the
individual responses combine positively. Substituting the angular
momentum variation given by Eq.~(\ref{deltaJ}) in Eq.~(\ref{magnetJ}),
and multiplying by a factor 2 to account for the positron
contribution, the induced dipole moment is
\begin{equation}
  \mathcal{M} = \frac{e^2r^2}{m}B\,,
  \label{magnetization}
\end{equation}
whence we get
\begin{equation}
  \tilde\mu_0= \frac{mr}{e^2}\,.
  \label{mu0ana}
\end{equation}

The expressions for $\tilde\varepsilon_0$ and $\tilde\mu_0$ must be
compatible with the Maxwell equations. Imposing
$\tilde\varepsilon_0\tilde\mu_0=1/c^2$, we find a relation between the
transition energy $\mathcal{E}_0$ and the radius $r$. Moreover, the
product $\tilde\varepsilon_0\tilde\mu_0$ does not depend on the volume
of the virtual pair since it cancels out in the ratio,
\begin{equation}
  c^2=\frac{1}{\tilde\varepsilon_0\,\tilde\mu_0}=\frac{\mathcal{M}/B}{\wp/E}\,,
  \label{ratioc}
\end{equation}
implying that in our model the speed of light is a consequence of the
magnetic and electric responses of each virtual pair locally.  Using
Eqs.~(\ref{dipmoment}) and~(\ref{magnetization}), The radius
associated with the volume of the virtual pair then reads
\begin{equation}
  r=\frac{c}{\omega_0}\,.
  \label{radius}
\end{equation}

Consistency with the Maxwell equations therefore results, by inserting
Eq.~(\ref{radius}) into Eq.~(\ref{varepsilonvalue}), in the following
expression for $\tilde\varepsilon_0$ (and
$\tilde\mu_0=1/\tilde\varepsilon_0c^2$),
\begin{equation}
  \tilde\varepsilon_0=\frac{\hbar\omega_0}{mc^2}\frac{e^2}{\hbar c}\,. 
  \label{epsilonmufinal}
\end{equation}
This equation can be written in terms of the fine structure constant
$\alpha=e^2/(4\pi\varepsilon_0\hbar c)$ and the known value of
$\varepsilon_0$ as\footnote{In Gaussian units, the fine structure
  constant is $\alpha=e^2/(\hbar c)$, and the result does not depend
  on the unit system.}
\begin{equation}
  \tilde\varepsilon_0=4\pi\alpha\frac{\hbar\omega_0}{mc^2}\varepsilon_0\,.
  \label{epsilonmufinal1}
\end{equation}

This expression shows that our simple model supplies numerical values
$\tilde\varepsilon_0$ and $\tilde\mu_0$ lying surprisingly close to
the correct values\footnote{Supposing $\hbar\omega_0=2mc^2$, one
  obtains the deviation factor
  $4\pi\alpha\hbar\omega_0/mc^2\approx1/10$. We think this is
  surprisingly close considering the crude semi-classical model
  employed.} $\varepsilon_0$ and $\mu_0$, in this manner supporting
their novel physical interpretation. Looking more closely, however,
one notices that Eq.~(\ref{epsilonmufinal}) tends to underestimate the
correct $\varepsilon_0$ and overestimate $\mu_0$ for sensible choices
of $\hbar\omega_0$. Also an astonishing property of
$\tilde\varepsilon_0$ and $\tilde\mu_0$ becomes apparent: they are
independent of the mass if, as expected, the energy gap depends
linearly on $m$. Indeed, one might be tempted to assume that only the
lightest elementary charged particles would contribute to the vacuum
polarization, since their virtual excitation would demand the smallest
amount of energy. But the independence of mass brings us to the
intriguing conclusion that all charged elementary particle pairs
should contribute more or less equally to the vacuum
response. Equation~(\ref{epsilonmufinal}) would then represent only
one partial contribution to the vacuum polarization. The total
response would be the sum of all partial
contributions~\cite{landau,sakharov}, but weighted by the square of
the charges $q_j$, where the index $j=1,2,\ldots$ stands for the
different virtual elementary particle
pairs. Equation~(\ref{epsilonmufinal}) is modified accordingly to
\begin{equation}
  \label{epsilonmufinalcharges1}
  \tilde\varepsilon_0^\mathrm{total}=4\pi\alpha\frac{\hbar\omega_0}{mc^2}\sum_j\left(\frac{q_j}{e}\right)^2\varepsilon_0\,.
\end{equation}
We note in passing that the speed of light is independent of the
number of particles, since the latter cancels out in
Eq.~(\ref{ratioc}). The equality between $\tilde\varepsilon_0$ and
$\varepsilon_0$ imposes a relation between the energy gap and the
number of elementary pairs, supplying an estimate for the latter,
\begin{equation}
  \sum_j\left(\frac{q_j}{e}\right)^2=\frac{1}{4\pi\alpha}\frac{mc^2}{\hbar\omega_0}\,.
  \label{totalcontrib}
\end{equation}
The choice $\hbar\omega_0=2mc^2$ would suggest around ten contributing
pairs with the electron charge.

In conclusion, our aim has been to demonstrate how a few reasonable
assumptions suffice to derive the correct orders of magnitude of
$\varepsilon_0$ and $\mu_0$, providing also a physical meaning to
these quantities\footnote{A first step in a more quantitative model
  would take into account the distinction between $r^2$ in
  Eq.~(\ref{magnetization}) and the average orbital radius needed for
  the diamagnetic term, here denoted by $\langle\rho^2\rangle$ (where
  $\rho$ is the distance to the axis in cylindrical coordinates),
  originating from the actual charge distribution. Assuming a constant
  probability density for the electron charge inside a solid spherical
  volume with radius $R$, an elementary calculation shows that
  $\langle\rho^2\rangle=2/5\times R^2$. Using this value in
  Eq.~(\ref{magnetization}) and the volume $V=4\pi/3\times R^3$ in
  Eq.~(\ref{varepsilonvalue}), we find $R=\sqrt{5/2}\times \hbar
  c/(\hbar\omega_0)$ by imposing
  $\tilde\varepsilon_0\tilde\mu_0=1/c^2$. These results change
  $\tilde\varepsilon_0$ of Eq.~(\ref{epsilonmufinal1}) to the new
  estimate
  \begin{equation}
    \tilde\varepsilon_0=3\alpha\left(\frac{2}{5}\right)^{3/2}\frac{\hbar\omega_0}{mc^2}\varepsilon_0\,.
  \end{equation}
  For the total number of virtual elementary pairs contributing to the
  vacuum response, one finds
  in place of Eq.~(\ref{totalcontrib}) the estimate
  \begin{equation}
    \sum_j\left(\frac{q_j}{e}\right)^2=\frac{1}{3\alpha}\left(\frac{5}{2}\right)^{3/2}\frac{mc^2}{\hbar\omega_0}\,,
  \end{equation}
  resulting in a number close to 90 pairs if
  $\hbar\omega_0=2mc^2$. Finally, we note that the actual charge
  distribution would be probably peaked around the origin, in this
  manner reducing the ratio $\langle\rho^2\rangle/R^2$ and increasing
  as a consequence the effective volume and the number of contributing
  virtual pairs.}.
Whilst regarded in classical electromagnetism as constants deprived of
deeper physical meaning, in the light of the quantum theory they would
be connected to fundamental physical processes, the polarization and
the magnetization of virtual pairs in vacuum in the linear regime
comprising weak and low-frequency (classical) fields. This alone we
think is worth noting. Additionally, one outcome of our model is an
independent estimate of the volume associated with a virtual
pair. Furthermore, our estimates also provide an evaluation of the
number of charged elementary particles. A more rigorous calculation
would have to rely on quantized fields~\cite{savolainenstenholm} and
on a potentially modified response closer to the resonance frequency.

In many of the seminal nonlinear effects of the quantum vacuum, the
contribution of the lightest elementary particles dominates. By
contrast, the linear effects discussed here enjoy equal contributions
from all elementary particles, including the ones not yet discovered,
owing to the independence of the masses of the virtual particles.

Finally, the association here investigated suggests how the Maxwell
equations would provide a connection between the value of the speed of
light and the magnitude of the linear response of vacuum predicted by
quantum physics. The situation is reminiscent of the decrease of the
speed of light caused by phase shifts inside a dielectric material. We
believe this to be a beautiful connection between, on the one hand,
the quantum concepts of virtual particles and vacuum fluctuations and,
on the other hand, the actual value of the maximum speed existent in
nature baring so many fundamental consequences. The fact that the
celebrated Lorentz invariant Maxwell equations would provide such a
bridge just adds to the surprise.

\begin{acknowledgement}

  We acknowledge enlightening discussions with Katiuscia N. Cassemiro,
  Pierre Chavel, Joseph H. Eberly, Michael Fleischhauer, Holger Gies,
  Uli Katz, Natalia V. Korolkova, G\'erard Mourou, Nicolai
  B. Narozhny, Serge Reynaud, Wolfgang P. Schleich, Anthony
  E. Siegman, and John Weiner.

\end{acknowledgement}



\end{document}